\documentclass[aip,jcp,showpacs,showkeys,reprint,floatfix]{revtex4-1}

\usepackage{graphicx}
\usepackage{color}
\usepackage[normalem]{ulem} 

\newcommand{\bq}{\begin{eqnarray}}
\newcommand{\eq}{\end{eqnarray}}
\newcommand{\bqn}{\begin{eqnarray*}}
\newcommand{\eqn}{\end{eqnarray*}}

\newcommand{\rr}{\mathbf{r}}
\newcommand{\RR}{\mathbf{R}}
\newcommand{\xx}{\mathbf{x}}
\newcommand{\yy}{\mathbf{y}}
\newcommand{\zz}{\mathbf{z}}

\newcommand{\qq}{\mathbf{q}}
\newcommand{\QQ}{\mathbf{Q}}

\newcommand{\calp}{{\cal P}}
\newcommand{\calh}{{\cal H}}
\newcommand{\calt}{{\cal T}}
\newcommand{\calv}{{\cal V}}
\newcommand{\cala}{{\cal A}}
\newcommand{\calo}{{\cal O}}
\newcommand{\cald}{{\cal D}}
\newcommand{\calr}{{\cal R}}

\begin{document}
\title{One-component fermion plasma on a sphere at finite temperature} 

\author{Riccardo Fantoni}
\email{rfantoni@ts.infn.it}
\affiliation{Universit\`a di Trieste, Dipartimento di Fisica, strada
  Costiera 11, 34151 Grignano (Trieste), Italy}
\date{\today}

\begin{abstract}
We study through a computer experiment, using the restricted path
integral Monte Carlo method, a one-component fermion plasma on a
sphere at finite, non-zero, temperature. We extract thermodynamic
properties like the kinetic and internal energy per particle and
structural properties like the radial distribution 
function. This study could be relevant for the characterization and
better understanding of the electronic properties of hollow graphene
spheres.   
\end{abstract}



\maketitle
\section{Introduction}
\label{sec:introduction}

We want to study the one-component fermion plasma on the surface of a 
sphere of radius $a$ at finite, non-zero, temperature, as an evolution
of the Thomson problem. The plasma is an ensemble of point-wise
electrons which interact through the Coulomb potential assuming that 
the electric field lines can permeate the tridimensional space where
the sphere is embedded. The system of particles is thermodynamically
stable even if the pair-potential is purely repulsive because the
particles are confined to the compact surface of the sphere, and we do
not need to add a uniform neutralizing background as in the Wigner
{\sl Jellium} model. Therefore our spherical plasma made of $N$ spinless
indistinguishable electrons of charge $-e$ and mass $m$ will carry a
total negative charge $-Ne$, a total mass $Nm$, and will have a radius
$a$.  

Note that in the limit $a\to\infty$ with a fixed surface density
$\sigma=N/4\pi a^2$ our system becomes thermodynamically unstable since
all the particles tends to escape to infinity. In order to prevent
this pathological scenario one would have to add a uniform neutralizing
background on the spherical surface of positive surface charge density
$+\sigma e$. This amounts to replacing the Coulomb potential $e^2/r$
with the corrected one $e^2/r-B$ with $B=\int_{\text{sphere}} (e^2/r) dA/(4\pi
a^2)=e^2$ where the integral is over the surface of the sphere
$dA=a^2\sin\theta\,d\theta d\varphi$ and $r=a\sqrt{2-2\cos\theta}$
is the Euclidean distance between the north pole and another point on
the sphere, with polar angle $\theta$. The 
constant $D$ is chosen to make sure that the average value of the
interaction is zero and must be subtracted from the self energy which
would otherwise be zero. We would then obtain the Wigner Jellium
system on the sphere which has received much attention from the point
of view of path integral Monte Carlo recently in the Euclidean
tridimensional space 
\cite{Brown2013,Brown2014,Dornheim2016,Dornheim2016b,Groth2016,Groth2017,Malone2016,Filinov2015}. 

We want to study the structural and thermodynamic properties at
finite, non-zero, temperature of the spherical fermion plasma through
restricted path integral Monte Carlo. In particular we will calculate
the radial distribution function of the particles on the surface of
the sphere and their kinetic and internal energy per particle.

Even if impenetrable identical particles on the surface of a sphere
admit a fractional anyonic statistics \cite{Lerda} we will just study
their fermionic nature, leaving the implementation of the anyonic
statistics to a subsequent work. This amounts to distinguish only among
even and odd permutations rather than among the larger elements of the
braid group. We will then consider the union of all the topologically
disjoint portion of the particles configuration space belonging just
to each of the two fermionic sections. This simplifies the problem
considerably since the braid group is much larger and complex than the
permutation group \cite{Lerda}. 

A quantum fluid on a Riemannian surface has been studied before
in relation to the quantum Hall effect
\cite{Prieto2009,Lee1999,Bergeron1996}. A generalized stochastic
method has also been implemented for the many-body ground state
\cite{Melik1997,Melik2001}. We are not aware of any path integral
Monte Carlo attempt in the spirit of our work. We expect our work
to be relevant for the characterization of the electronic properties
of hollow graphene spheres \cite{Rashid,Tiwari} constructed in the
laboratory and for their implementation as electrodes for
supercapacitors and batteries, as superparamagnetic materials,
electrocatalysts for oxygen reduction, as drug deliverers, as a 
conductive catalyst for photovoltaic applications
\cite{Guo2010,Cao2013,Wu2013,Shao2013,Zhao2016,Cho2016,Hao2016,Huang2017,Chen2017}. 
Our numerical experiments albeit   
idealized are capable to explore the properties of these systems under
the most various thermodynamic conditions, even extreme conditions
otherwise not accessible in the laboratory. Therefore we are able to
explore and characterize the phenomenology of these systems with
cost-free computer experiments that can later be used as guides for
the laboratory set up.
 
The paper is organized as follows: in section \ref{sec:problem} we
describe the problem we want to solve and the method used for its
resolution, in section \ref{sec:results} we present our numerical
results, and section \ref{sec:conclusions} is for concluding
discussion. 

\section{The problem}
\label{sec:problem}

A point $\qq$ on the sphere of radius $a$, the surface of constant
positive curvature, is given by 
\bq
\rr/a=\sin\theta\cos\varphi\hat{\xx}+\sin\theta\sin\varphi\hat{\yy}+
\cos\theta\hat{\zz},
\eq
with $\theta$ the polar angle and $\varphi$ the azimuthal angle.
The $N$ particles positions are at
$\RR=(\rr_1,\rr_2,\ldots,\rr_N)$. The surface density of the plasma
will then be $\sigma=N/4\pi a^2$.
On the sphere we have the following metric
\bq
ds^2=g_{\mu\nu}dq^\mu dq^\nu=a^2\left[d\theta^2+\sin^2\theta d\varphi^2\right],
\eq 
where Einstein summation convention on repeated indices is assumed, we
will use Greek indices for either the surface components or the surface
components of each particle coordinate and roman indices for either
the particle index or the time-slice index, $q^1=\theta\in
[0,\pi)$, $q^2=\varphi\in [-\pi,\pi)$, and the positive definite
and symmetric metric tensor is given by
\bq
g_{\mu\nu}=\left(\begin{array}{cc}
a^2 & 0 \\
0   & a^2\sin^2\theta
\end{array}\right).
\eq
We have periodic boundary conditions in $\theta+\pi=\theta$ and in
$\varphi+2\pi=\varphi$. We will not need 
to implement explicitly the periodic boundary conditions as all that
is needed in the simulation is the geodesic and the Euclidean distance
which are expressed in terms of trigonometric functions which are
periodic in the coordinates $\theta$ and $\varphi$. We will also define 
$\QQ=(\qq_1,\qq_2,\ldots,\qq_N)$ which will be the coordinates used in
the code. The geodesic distance between two infinitesimally close
points $\QQ$ and $\QQ'$ is
$ds^2(\QQ,\QQ')=\sum_{i=1}^Nds^2(\qq_i,\qq_i')$ where the geodesic
distance between the points $\qq$ and $\qq'$ on the sphere is
\bq \label{gd}
s(\qq,\qq')&=&a\arccos\left[\cos(q^1)\cos({q^1}')+\right.\\
&&\left.\sin(q^1)\sin({q^1}')\cos(q^2-{q^2}')\right],
\eq
On a computer the haversine formula is numerically better conditioned
for small distances. Moreover, to avoid rounding errors for the special
case of antipodal points the Vincenty formula for an ellipsoid with
equal major and minor axes may be used.

The Hamiltonian of the $N$ non-relativistic indistinguishable particles
of the one-component spinless fermion plasma is given by 
\bq \label{Ham}
\calh=\calt+\calv=-\lambda\sum_{i=1}^N\Delta_i+\sum_{i<j}v_{ij},
\eq
with $\lambda=\hbar^2/2m$, where $m$ is the electron mass, and
$\Delta_i=g_i^{-1/2}\partial(g_i^{1/2} g_i^{\mu\nu}\partial/\partial
q_i^\nu)/\partial q_i^\mu$ the Laplace-Beltrami operator for
the $i$th particle on the sphere of radius $a$ in local coordinates,
where $g_{\mu\alpha}g^{\alpha\nu}=\delta_\mu^\nu$ and $g_i=\det
||g_{\mu\nu}(\qq_i)||$. We have assumed that 
$\calh$ in curved space has the same form as in flat space. For the
pair-potential, $v$, we will choose 
\bq
v_{ij}=e^2/r_{ij},
\eq
where $e$ is the electron charge and $r_{ij}$ is the Euclidean
distance between two particles at $\qq_i$ and $\qq_j$, which is given
by 
\bq
r_{ij}=a\sqrt{2-2\hat{\rr}_i\cdot\hat{\rr}_j}=
2a\sin[\arccos(\hat{\rr}_i\cdot\hat{\rr}_j)/2],
\eq
where $\hat{\rr}_i=\rr_i/a$ is the versor that from the center of the
sphere points towards the center of the $i$th particle. 

Given the antisymmetrization operator $\cala$, and the inverse
temperature $\beta=1/k_BT$, with $k_B$ Boltzmann's constant, the
one-component fermion plasma density matrix, $\rho_F=\cala
e^{-\beta\calh}$, in the coordinate representation, on a generic
Riemannian manifold of metric $g$ \cite{Fantoni12e,Schulman}, is
\begin{widetext}
\bq \nonumber
\rho_F({\QQ^\prime},\QQ;\beta)=\int&&
\rho_F({\QQ^\prime},\QQ((M-1)\tau);\tau)
\cdots
\rho_F(\QQ(\tau),\QQ;\tau)\times\\ \label{rhof}
&&\prod_{j=1}^{M-1}\sqrt{\tilde{g}_{(j)}}\prod_{i=1}^N\,
dq_i^1(j\tau)\wedge dq_i^2(j\tau)~,
\eq
where as usual we discretize the {\sl imaginary thermal time} in bits
$\tau=\hbar\beta/M$. We will often use the following shorthand
notation for the {\sl path integral} measure:
$\prod_{j=1}^{M-1}\sqrt{\tilde{g}_{(j)}}\prod_{i=1}^N\,$ 
$dq_i^1(j\tau)\wedge dq_i^2(j\tau)\to\cald\QQ$ as $M\to\infty$. The
path of the $i$th particle is given by
$\{\qq_i(t)|t\in[0,\hbar\beta]\}$ with $t$ the 
imaginary thermal time. Each $\qq_i(j\tau)$ with $i=1,\ldots,N$ and
$j=1,\ldots,M$ represents the varius {\sl beads} forming the
discretized path. The $N$ particle path is given by
$\{\QQ(t)|t\in[0,\hbar\beta]\}$. Moreover,  
\bq
\tilde{g}_{(j)}&=&\det||\tilde{g}_{\mu\nu}(\QQ(j\tau))||,~~~j=1,2,\ldots,M-1,\\
\tilde{g}_{\mu\nu}(\QQ)&=&g_{\alpha_1\beta_1}(\qq_1)\otimes\ldots
\otimes g_{\alpha_N\beta_N}(\qq_N),
\eq
In the small $\tau$ limit we have 
\bq \nonumber
\rho_F(\QQ(2\tau),\QQ(\tau);\tau)=(2\pi\hbar)^{-N}\cala\left[
\tilde{g}_{(2)}^{-1/4}\sqrt{D(\QQ(2\tau),\QQ(\tau);\tau)}
\tilde{g}_{(1)}^{-1/4}\times\right.\\
\left.e^{\lambda\tau R(\QQ(\tau))/6\hbar}
e^{-\frac{1}{\hbar}S(\QQ(2\tau),\QQ(\tau);\tau)}\right],
\eq
\end{widetext}
where $\cala$ can act on the first, or on the second, or on both {\sl
time slices}, $R(\QQ)$ the scalar curvature of the curved manifold,
$S$ the action and $D$ the van Vleck's determinant  
\bq
D_{\mu\nu}&=&-\frac{\partial^2S(\QQ(2\tau),\QQ(\tau);\tau)}
{\partial {Q}^\mu(2\tau)\partial {Q}^\nu(\tau)},\\
\det||D_{\mu\nu}||&=&D(\QQ(2\tau),\QQ(\tau);\tau),
\eq
where here the Greek index denotes the two components of each particle
coordinate.  

For the {\sl action} and the {\sl kinetic-action} we have 
\bq
S(\QQ',\QQ)&=&K(\QQ',\QQ)+U(\QQ',\QQ),\\
K(\QQ',\QQ)&=&\frac{3N\hbar}{2}\ln(4\pi\lambda\tau/\hbar)+
\frac{\hbar^2s^2(\QQ',\QQ)}{4\lambda\tau},
\eq
where in the {\sl primitive approximation} \cite{Ceperley1995} we find
the following expression for the {\sl inter-action}, 
\bq \label{primitive}
U(\QQ',\QQ)&=&\frac{\tau}{2}[V(\QQ')+V(\QQ)],\\
V(\QQ)&=&\sum_{i<j}v_{ij}.
\eq
In particular the kinetic-action is responsible for a diffusion of the
random walk with a variance of $2\lambda\tau g^{\mu\nu}/\hbar$.

On the sphere we have $R=N\calr$ with $\calr=2/a^2$, the scalar
curvature of the sphere of radius $a$, and in the $M\to\infty$ limit 
$s(\QQ',\QQ)\to ds(\QQ',\QQ)$ and
$\tilde{g}_{(2)}^{-1/4}$ $\sqrt{D(\QQ(2\tau),\QQ(\tau);\tau)}$ 
$\tilde{g}_{(1)}^{-1/4}\to\left(\hbar^2/2\lambda\tau\right)^{N}$. We
recover the Feynman-Kac path integral formula on the sphere in the
$\tau\to 0$ limit. In a computer experiment calculation it is enough to
take $M$ sufficiently large, of the order of 100 or 1000
\cite{Ceperley1995}, so to keep $\tau\sim 0.01$, recalling that the
primitive approximation error 
scales as $\sim\lambda\tau^2$. We will then have to deal with $2NM$
multidimensional integrals for which Monte Carlo is a suitable
computational method. For example to measure an observable $\calo$ we
need to calculate the following quantity
\bq
\langle\calo\rangle=\frac{\int
  O(\QQ,\QQ')\rho_F(\QQ',\QQ;\beta)\,d\QQ d\QQ'}
{\int \rho_F(\QQ,\QQ;\beta)\,d\QQ},
\eq
where $\sqrt{\tilde{g}}\prod_{i=1}^N\,dq_i^1\wedge dq_i^2\equiv
d\QQ$. Notice that most of the properties 
that we will measure are diagonal in coordinate representation,
requiring then just the diagonal density matrix,
$\rho_F(\QQ,\QQ;\beta)$. For example for the radial distribution
function, $g(r)=\langle\calo\rangle$ with $r$ the Euclidean distance
between points $\qq$ and $\qq'$,
$r=2a\sin[\arccos(\hat{\qq}\cdot\hat{\qq}')/2]$, we have the following
histogram estimator,   
\bq
O(\QQ;r)=\sum_{i\neq j}
\frac{1_{[r-\Delta/2,r+\Delta/2[}(q_{ij})}{Nn_{id}(r)},
\eq
where $\Delta$ is the histogram bin, $1_{[a,b[}(x)=1$ if $x\in[a,b[$
and 0 otherwise, and 
\bq
n_{id}(r)=N\left[\left(\frac{r+\Delta/2}{2a}\right)^2-
\left(\frac{r-\Delta/2}{2a}\right)^2\right]~,
\eq
is the average number of particles on the spherical crown
$[r-\Delta/2,r+\Delta/2[$ for the ideal gas of density
$\sigma$. We have that $\sigma^2g(r)$ gives the probability that
sitting on a particle at $\qq$ one has to find another particle at
$\qq^\prime$.  

Fermions' properties cannot be calculated exactly with path integral
Monte Carlo because of the fermions sign problem
\cite{Ceperley1991,Ceperley1996}. We then have to resort to an
approximated calculation. The one we chose was the restricted path
integral approximation \cite{Ceperley1991,Ceperley1996} with a ``free
fermions restriction''. The trial density matrix used in the
restriction is chosen as the one reducing to the ideal density matrix
in the limit of $t\ll 1$ and is given by
\bq \label{ifdm}
\rho_0(\QQ',\QQ;t)\propto\cala \left|\left|e^{
-\frac{\hbar s^2(\qq_i',\qq_j)}
{4\lambda t}}\right|\right|.
\eq
The {\sl restricted path integral identity} that we will use states
\cite{Ceperley1991,Ceperley1996} 
\bq \label{rpii}
\rho_F(\QQ',\QQ;\beta)&\propto&\int \sqrt{\tilde{g}''}d\QQ''\,\rho_F(\QQ'',\QQ;0)\times
\\ \nonumber
&&\oint_{\QQ''\to\QQ'\in\gamma_0(\QQ)}\cald\QQ'''\,e^{-S[\QQ''']/\hbar},
\eq
where $S$ is the Feynman-Kac action
\bq
S[\QQ]=\int_0^{\hbar\beta} dt\left[\frac{\hbar^2}{4\lambda}
\dot{\QQ}_\mu\dot{\QQ}^\mu+V(\QQ)\right],
\eq
here the dot indicates a total derivative with respect to the
imaginary thermal time, and the subscript in the path integral of
Eq. (\ref{rpii}) means that we restrict the path integration to 
paths starting at $\QQ''$, ending at $\QQ'$ and avoiding the nodes of
$\rho_0$, that is to the {\sl reach} of $\QQ$, $\gamma_0$. The nodes
are on the reach boundary $\partial\gamma_0$. The weight of
the walk is $\rho_F(\QQ'',\QQ;0)=\cala\delta(\QQ''-\QQ)$
$=(N!)^{-1}\sum_\calp(-)^\calp$ $\delta(\QQ''-\calp\QQ)$, where the
sum is over all the permutations $\calp$ of the $N$ fermions,
$(-)^\calp$ is the permutation sign, positive for an even permutation
and negative for an odd permutation, and the Dirac's delta function is
on the sphere. It is clear that the  
contribution of all the paths for a single element of the density
matrix will be of the same sign, thus solving the sign problem;
positive if $\rho_F(\QQ'',\QQ;0)>0$, negative otherwise. On the
diagonal the density matrix is positive and on the path restriction
$\rho_F(\QQ',\QQ;\beta)>0$ then only even permutations are allowed
since $\rho_F(\QQ,\calp\QQ;\beta)=(-)^\calp\rho_F(\QQ,\QQ;\beta)$. It
is then possible to use a bosons calculation to get the fermions
case. Clearly the restricted path integral identity with the free
fermions restriction becomes exact if we simulate free fermions, but
otherwise is just an approximation. The approximation is expected to
become better at low density and high temperature, i.e. when
correlation effects are weak. The implementation of the restricted,
fixed nodes, path integral identity within the worm algorithm has been
the subject of a recent study on the tridimensional Euclidean Jellium.

We will use the {\sl worm algorithm}
\cite{Prokofev1998,Boninsegni2006a} to generate spontaneously the
needed permutations for the antisymmetrization operator $\cala$. The
permutations on the sphere will generate paths with different braiding
properties. Identical impenetrable (scalar) particles on a sphere are,
in general, anyons with fractional statistics \cite{Lerda}. Here we
will just project out the fermionic component of the broader braid
group by just looking at the sign of the trial free fermions density
matrix. It is still object of study the realization of the simulation
of the anyonic system. The worm algorithm is able to sample the
necessary permutations of the indistinguishable particles without the
need of explicitly sampling the permutations space 
treating the paths as ``worms'' with a tail ({\sl Masha}) and a head
({\sl Ira}) in the $\beta$-periodic imaginary time, which can be
attached one with the other in different ways or swap some of their
portions. 

We will work in the grand canonical ensemble with fixed chemical
potential $\mu$, surface area $A=4\pi a^2$, and absolute temperature
$T$. At a higher value of the chemical potential we will have a higher
number of particles on the surface and a higher density. On the other
hand, increasing the radius of the sphere at constant chemical
potential will produce a plasma with lower surface density.  
The {\sl Coulomb coupling constant} is $\Gamma=\beta e^2/a_0r_s$ with
$a_0=\hbar^2/me^2$ the Bohr radius and
$r_s=(4\pi\sigma)^{-1/2}/a_0$. At weak coupling, $\Gamma\ll 1$, the 
plasma becomes weakly correlated and approach the ideal gas
limit. This will occur at high temperature and/or low density. The
{\sl electron degeneracy parameter} is $\Theta=T/T_D$ where the
degeneracy temperature $T_D=\sigma\hbar^2/mk_B$. For temperatures
higher than $T_D$, $\Theta\gg 1$, quantum effects are less relevant.

\section{Results}
\label{sec:results}

Choosing length in Wigner-Seitz's radius, $a_0r_s$, units and energies
in Rydberg's, $\text{Ry}=\hbar^2/2ma_0^2$, units we have
$\lambda=\text{Ry}/r_s^2$, $\Gamma=\beta(2/r_s)$, and
$\Theta=(2\pi r_s^2)/\beta$. We then see immediately that when quantum
effects are relevant, at $\Theta\lesssim 1$, and at low density or high 
$r_s$ the potential energy dominates in the Hamiltonian (\ref{Ham})
and the electron plasma tends to crystallize in a Wigner's
crystal. On the other hand at $\Theta\gg 1$, in the classical regime,
the system tends to crystallize at high density. In our grand
canonical simulation is rather convenient to choose the length unit to
be just the Bohr radius since $r_s$ is not an input parameter. 

We use a free fermion trial density matrix restriction for the fixed
nodes path integral calculation from the worm algorithm
\cite{Boninsegni2006a,Boninsegni2006b} to the reach of the reference
point in moves ending in the Z sector: remove, close, wiggle, and 
displace. We will use the primitive approximation of
Eq. (\ref{primitive}). Our algorithm has been recently described in
Ref. \cite{Fantoni2018a}. Here we do not randomize the reference point
time slice and we do not restrict the G sector. We
choose the probability of being in the G sector ($\propto C_0$ in Ref.
\cite{Boninsegni2006a}) so as to have Z sector's acceptance ratio
close to 8/10. The restriction implementation is rather  
simple: we just reject the move whenever the proposed path is such
that the ideal fermion density matrix (\ref{ifdm}) calculated between
the reference point and any of the time slices subject to newly
generated particles positions has a negative value. The algorithm will
spontaneously choose the optimal needed $\tau$, in the sense that for
bigger $\tau$ it will not be able to come back and forth between the Z
and the G sector remaining stuck in the G sector. 

\begin{table*}[htbp]
\protect\caption{Thermodynamic states treated in our simulations:
  $\mu~(\text{Ry})$ chemical potential, $\beta~(\text{Ry}^{-1})$
  inverse temperature, $\overline{N}$ average number of particles,
  $\overline{r_s}$ average value of $r_s$, $e_K~(\text{Ry})$ kinetic
  energy per particle from the thermodynamic estimator as exaplained
  in Ref. \cite{Ceperley1995}, and $e_V~(\text{Ry})$ 
  potential energy per particle. The other quantities were
  introduced in the main text. We chose length in Bohr
  radius' units and energy in Rydberg's units. We chose $M$ such as to
  have $\tau=0.01$ or less in all cases except case A where we have
  $\tau=0.02$.}  
\vspace{1cm}
\label{tab:tq}
\begin{ruledtabular}
\begin{tabular}{|c|llllllllll|}
case & $M$ & $\mu$ & $a/a_0$ & $\beta$ & $\overline{N}$ &
$\overline{r_s}$ & $\Gamma$ & $\Theta$ & $e_K$ & $e_V$ \\ 
\hline
A & 2000 & 4 & 5 & 40 & 15.03(3) & 1.29 & 62.0 & 0.261 &
24.9(3) & 2.67(3) \\
B & 500 & 8 & 5 & 5 & 20.80(8) & 1.10 & 9.12 & 1.51 &
48.97(4) & 3.857(6) \\
C & 100 & 10 & 5 & 1 & 29.2(2) & 0.925 & 2.16 & 5.38 &
48.5(2) & 6.08(5) \\
D & 50 & 8 & 5 & 1/2 & 31.0(1) & 0.898 & 1.11 & 10.1 &
47.84(6) & 6.43(3) \\
E & 10 & $-$13 & 5 & 1/10 & 61.8(3) & 0.636 & 0.314 & 25.4 &
51(1) & 12.83(6) \\
F & 2 & $-$300 & 5 & 2/100 & 58.9(1) & 0.651 & $6.14\times
10^{-2}$ & 133 & 61(4) & 11.86(2) \\
G & 2 & $-$250 & 5 & 0.015 & 48.00(3) & 0.722 & $4.16\times
10^{-2}$ & 218 & $-$9(4) & 9.412(5) \\
\hline
H & 100 & 4 & 10 & 1 & 35.3(2) & 1.68 & 1.19 & 17.8 &
$-$38(36) & 3.90(2) \\
I & 100 & 0 & 20 & 1 & 50.5(4) & 2.81 & 0.711 & 49.8 &
42(3) & 3.02(3) \\
L & 100 & $-$8 & 200 & 1 & 17.7(2) & 47.5 & $4.21\times
10^{-2}$ & $1.42\times 10^{4}$ & 45(3) & 0.118(1) \\
\end{tabular}
\end{ruledtabular}
\end{table*}

The restricted worm algorithm simulations length was $n\times 10^3$
blocks, with $n\in [0,10]$ an integer. Each block was made of 500
steps during 
which 100 moves where made and measures and averages taken. The moves
were of 9 kinds: advance, recede, insert, open, and swap ending in the
G sector; remove, close, wiggle, and displace ending in the Z sector
\cite{Fantoni14d}. Each move involved no more than 20 time slices. And
they were chosen from a menu with equal probabilities. The integration
measures factors $\sqrt{\tilde{g}}$ where only used in the acceptance
probabilities of the self-complementary moves: wiggle, swap, and
displace. 

In table \ref{tab:tq} we show the cases studied in our
simulations. The first case A is at a temperature of about $3946$~K
below the graphene melting temperature \cite{Los2015}. From the table
we see how the potential energy per particle diminishes as the density
of the system decreases. 

\begin{figure}[htbp]
\begin{center}
\includegraphics[width=8cm]{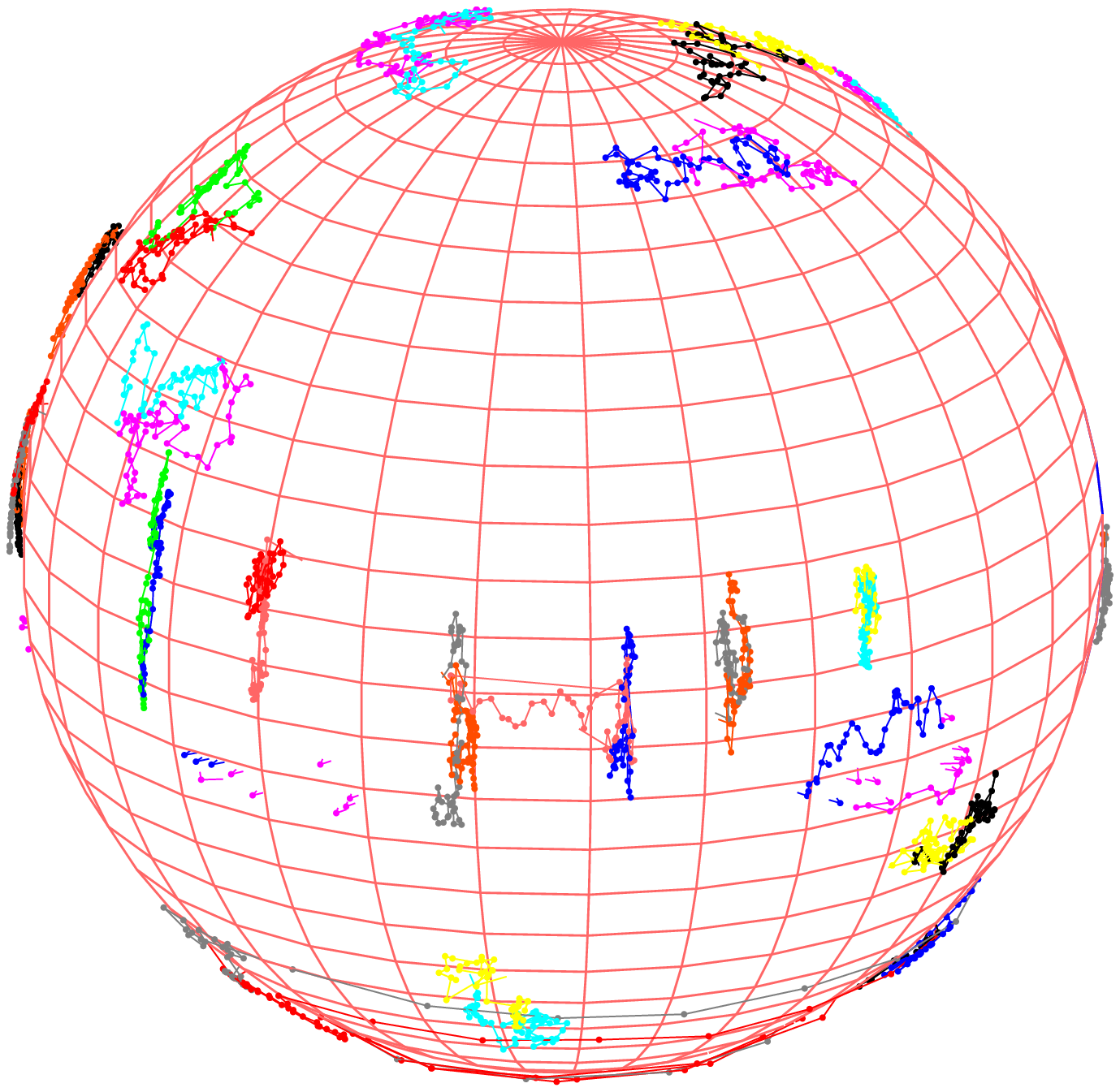}
\includegraphics[width=8cm]{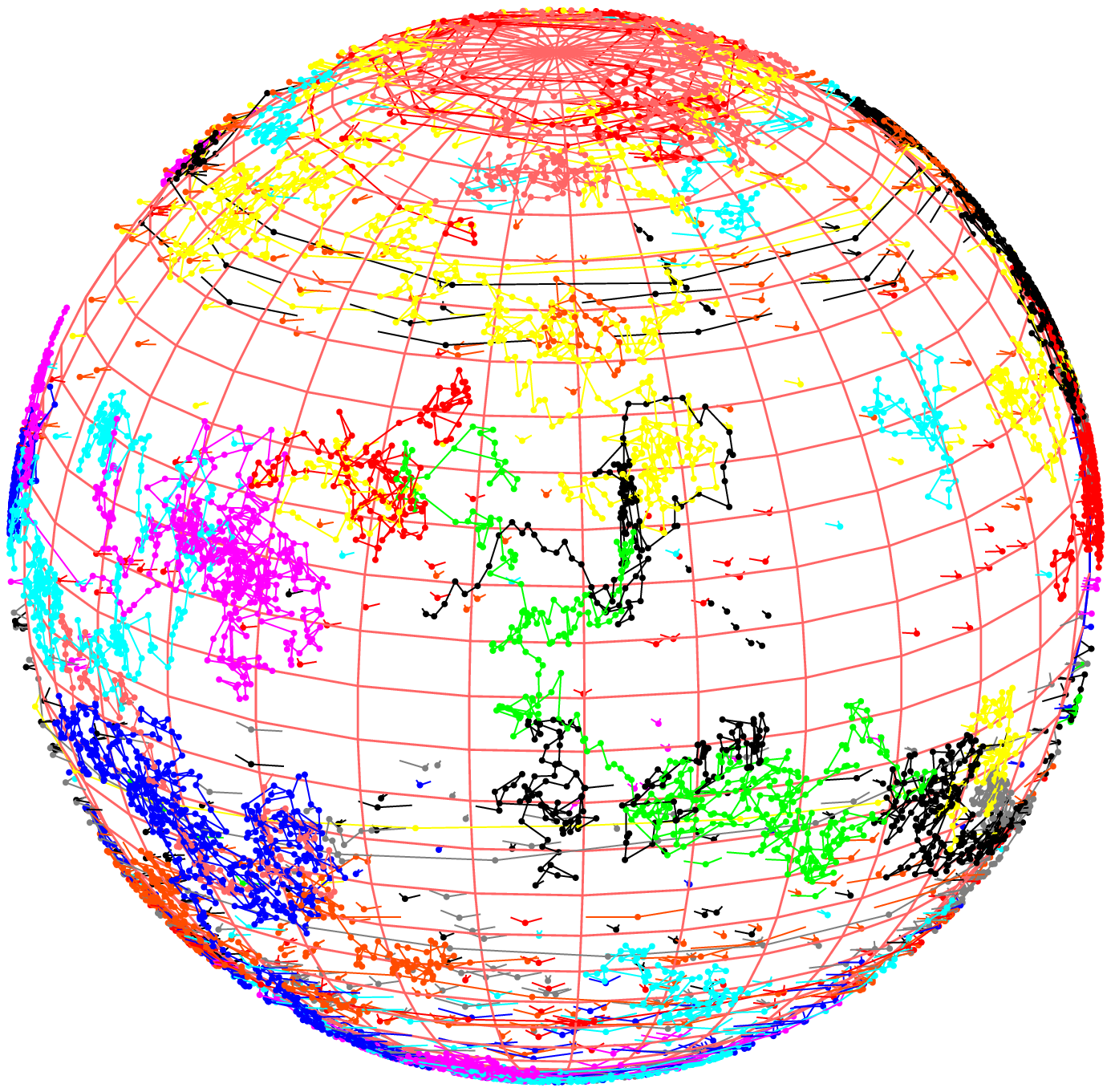}
\end{center}  
\caption{(color online) Snapshot of the macroscopic path during the
  simulation of case C in table \ref{tab:tq} in the top panel and case
  B in table \ref{tab:tq} in the bottom panel. The different worms
  have different colors. Some paths penetrate through the surface of
  the sphere and appear as broken links.}
\label{fig:path}
\end{figure}

In figure \ref{fig:path} we show a snapshot of the macroscopic path
during an equilibrated simulation of case B and C of table
\ref{tab:tq}. We see how the particles tend to cover the sphere surface
isotropically. As it should be since there is nothing able to break
the symmetry. Regarding the paths configuration we see immediately
that the ones in case B, at lower temperature, are more extended than
the ones in case C, at higher temperature, in agreement with the fact
that the de Broglie thermal wavelength, the size of a path in absence
of interactions, is bigger in case B. We can distinguish between 
several kinds of conformations. There are the localized paths and the
unlocalized path covering a large portion of the sphere surface. Paths
tend to avoid the poles at low temperature. They tend to wind around
the sphere running along the parallels in proximity of the poles and to
run along the meridians in proximity of the equator. This is because
these are the paths favored by the kinetic-action which is expressed
in terms of the square of the geodesic distance of 
Eq. (\ref{gd}) which, unlike the euclidean distance, is homogeneous
only in the azimuthal angle, the $q^2$ local coordinate, but not in
the polar angle, the $q^1$ local coordinate. At lower temperature,
when the path size increases, the worm diffuses more and we can have
paths covering a larger part of the sphere with longer links between
two beads. If we rotate the sphere moving its $\hat{\zz}$ axis the
paths configuration will also rotate. 

\begin{figure}[htbp]
\begin{center}
\includegraphics[width=8cm]{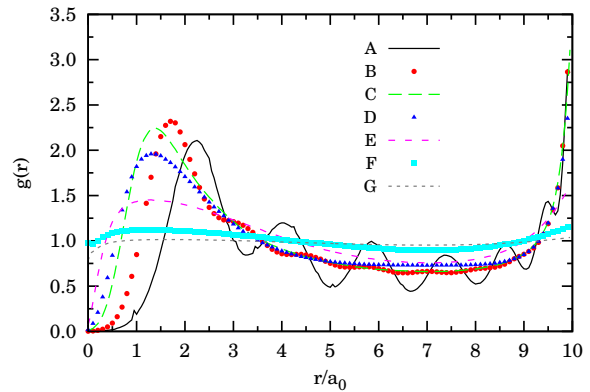}
\end{center}  
\caption{(color online) The radial distribution function for the
  spinless fermion 
  plasma on the sphere of radius $a=5a_0$ at an inverse temperature
  $\beta$ and a chemical potential $\mu$ for the cases A-G shown in
  table \ref{tab:tq}.}    
\label{fig:gr}
\end{figure}

In figure \ref{fig:gr} we show the radial distribution function for
the cases shown in table \ref{tab:tq}. Note that here we are plotting
against the Euclidean distance instead of the geodesic one so the
value of $g(r)$ on the diameter is at $r=a\sqrt 2$, the on top value is
at $r=0$, and the antipodal value is at $r=2a$. We then see the effect
of curvature on the Coulomb and Fermi hole near contact as they evolve
by increasing the temperature. The extent of the Coulomb and Fermi
hole at the lowest temperature amounts to roughly 2.5 Bohr's
radii. In the limit of very high temperature the radial
distribution function tends to the constant function everywhere equal
to unity (see case G of table \ref{tab:tq}).  
Another feature of the radial distribution function is the first peak
which is produced due to the Pauli exclusion principle, responsible of
the Fermi hole, to the Coulomb repulsion, responsible of the Coulomb
hole, and to the temperature effect which tends to make particles bump
one on the other. From the figure we clearly see how at small
$\Theta$, when the Pauli exclusion becomes strong, the peak tends to
shift at larger distances. At very high $\Theta$, the Pauli exclusion
becomes very weak and the Fermi hole tends to disappear. Curiously
enough the height of the first peak, the probability that sitting
on a particle we find one in its neighborhood, is lower than the
antipodal value, probability of finding a particle to the particle
antipodes. The first peak height and the antipodal value have a non
monotonic behavior with temperature. Since there are no
attractions in the pair-potential we only observe oscillations in the
radial distribution function at very low temperature.

\begin{figure}[htbp]
\begin{center}
\includegraphics[width=8cm]{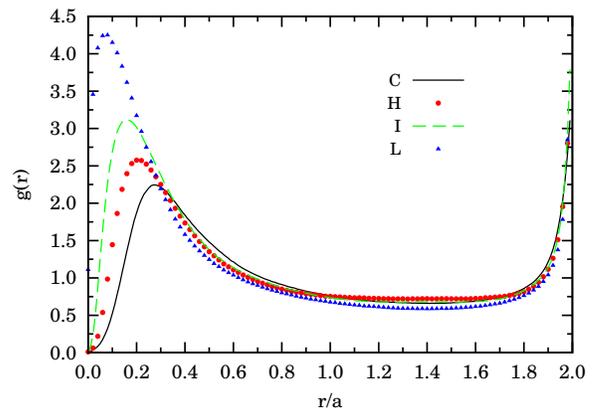}
\end{center}  
\caption{(color online) The radial distribution function for the
  spinless fermion 
  plasma on the sphere of different radii at an inverse temperature
  $\beta=1~\text{Ry}^{-1}$ and a chemical potential $\mu$ for the
  cases shown in table 
  \ref{tab:tq}.}   
\label{fig:gr-density}
\end{figure}

In figure \ref{fig:gr-density} we show the radial distribution
function of the plasma at the inverse temperature
$\beta=1~\text{Ry}^{-1}$ on spheres of different diameters and with
roughly equal average number of particles, as shown in table
\ref{tab:tq} for cases C, H, I, and L. Case L corresponds to a sphere
of the diameter of 20 nanometers and still presents the Coulomb and
Fermi hole. We can see that, as the diameter increases and the density
decreases, the first peak height increases. This had to be expected in
view of the fact that the system in the semi-quantal regime will tend
to crystallize as the density decreases. The peak height tends to
become bigger than the antipotal value. 

We always worked with no more than 65 electrons which could
corresponds to the $\pi$ conduction electrons of the carbon atoms in
the graphene sphere. So the spheres should be made by 10-100 C
atoms. The same order of magnitude as in fullerenes where the smallest
buckyball cluster is C$_{20}$ and the most common is the
buckminsterfullerene C$_{60}$. Here we are not taking care of the fact
that, in graphene, at the Dirac point, electrons have zero effective
mass. These graphinos should have a relativistic Hamiltonian rather
than the non-relativistic one we used in Eq. (\ref{Ham}). 

\section{Conclusions}
\label{sec:conclusions}

We simulated a one-component spinless fermion plasma at finite,
non-zero, temperature on the surface of a sphere. The Coulomb
interaction is $e^2/r$ with $r$ the Euclidean distance between the two
electrons of elementary charge $e$. Here we could as well have chosen
instead of $r$ the geodesic distance, $s$, within the sphere. We used
a new implementation of the restricted fixed nodes path integral
identity within the worm Monte Carlo algorithm. This gives 
us an approximated numerical solution of the many-body problem. The
exact solution cannot be accessed due to the fermion sign
catastrophe. impenetrable indistinguishable particles on the surface
of a sphere admit, in general, anyonic statistics. Here we just project
the larger bride group onto the permutation group and choose the
fermion sector for our study.

The path integral Monte Carlo method chosen uses the primitive
approximation for the action which could be improved for example by
the use of the pair-product action \cite{Ceperley1995}. The
restriction is carried on choosing as the trial density matrix the one
of ideal free fermion. This choice would return an exact solution for
the simulation of ideal fermions but it furnishes just an approximation
for the interacting coulombic plasma.

Our results extend to the quantum regime the previous non-quantum
results obtained for the analytically exactly solvable plasma on
curved surfaces 
\cite{Fantoni03jsp,Fantoni2008,Fantoni2012,Fantoni2012b,Fantoni2016,Fantoni17b}
and for its numerical Monte Carlo experiment \cite{Fantoni12c}. Here
we just study the geometry of the sphere leaving the more complex
surfaces with a non-constant curvature to a further study. As is shown
by the snapshot of the macroscopic path, the configuration space
appears much more complicated than in the classical case (see Figs. 5
and 6 of Ref. \cite{Fantoni12c}). A first notable phenomena is the
fact that whereas the particles distribution is certainly isotropic
the paths conformation is not, with beads distributed in such way to
avoid the poles at low temperature. Some paths tend to wind around
the sphere running along the parallels in proximity of the poles
others to run along the meridians in proximity of the equator. This is
a direct consequence 
of the coordinate dependence of the variance of the diffusion. If we
rotate the sphere the path configuration will also roatate with the
sphere. We have several kinds of worms conformations. At high and low
temperature: The localized ones, those winding around the sphere along
parallels, and those penetrating through the surface of the sphere. At
low temperature the unlocalized ones distributed over a larger part of
the surface with long links between the beads of the path.

The structure of the plasma on the sphere reveals how the curvature
influences the Coulomb and Fermi holes as they evolve in temperature
and density. In particular we observe a monotonic increase of the
extent of the Fermi hole as the temperature diminishes. Our analysis
shows how the probability of finding a 
particle nearby another particle is lower than the probability of
finding a particle at the antipodes unless for spheres of large
diameter. At a higher degeneracy parameter the Pauli exclusion effect
becomes less important and the Fermi hole tends to disappear. In the
high temperature limit the particles will tend to cover the sphere
more uniformly. Decreasing the surface density at fixed low
temperature the first peak of the radial distribution function grows
monotonically in height, tends to become bigger than its antipodal
value, and shifts at smaller distances.

Our computer experiment could be used to predict the properties of a
metallic spherical shell, as for example a spherical shell of
graphene. Today we assisted to the rapid development of the laboratory
realization of graphene hollow spheres with many technological
interests like the employment as electrodes for supercapacitors and
batteries, as superparamagnetic materials, as electrocatalysts for
oxygen reduction, as drug deliverers, as a conductive catalyst for
photovoltaic applications. Of course, with simulation we can access the
more various and extreme conditions otherwise not accessible in a
laboratory. 

A possible further study would be the simulation of the neutral sphere
where we model the plasma of electrons as embedded in a spherical
shell that is uniformly positively charged in such a way that the
system is globally neutrally charged. This can easily be done by
changing the Coulomb pair-potential into $e^2/r\to e^2(1/r-1)$. In the
$a\to\infty$ limit, this would reduce to the Wigner Jellium model
which has been received much attention lately, from the point of view
of a path integral Monte Carlo simulation
\cite{Brown2013,Brown2014,Dornheim2016,Dornheim2016b,Groth2016,Groth2017,Malone2016,Filinov2015}. 
Alternatively
we could study the two-component plasma on the sphere as has recently
been done in the tridimensional Euclidean space \cite{Fantoni2018b}. 
Another 
possible extension of our work is the realization of the simulation of
the full anyonic plasma on the sphere taking care appropriately of the
fractional statistics and the phase factors to append to each
disconnected region of the path integral expression for the partition
function. 
This could become important in a study
of the quantum Hall effect by placing a magnetic Dirac monopole at the
center of the sphere \cite{Melik1997,Melik2001}. Also the adaptation
of our study to a fully relativistic Hamiltonian could be of some
interest for the treatment of the Dirac points graphinos.

\begin{acknowledgments}
I would like to dedicate this work to my daughter Alice and to my wife
Laure without whose support all this could not have been
accomplished. I would like to acknowledge relevant discussions with
Saverio Moroni whose comments and suggestions have been invaluable
through the whole creation of the work. 
\end{acknowledgments} 
\bibliography{jft}

\begin{thebibliography}{44}%
\makeatletter
\providecommand \@ifxundefined [1]{%
 \@ifx{#1\undefined}
}%
\providecommand \@ifnum [1]{%
 \ifnum #1\expandafter \@firstoftwo
 \else \expandafter \@secondoftwo
 \fi
}%
\providecommand \@ifx [1]{%
 \ifx #1\expandafter \@firstoftwo
 \else \expandafter \@secondoftwo
 \fi
}%
\providecommand \natexlab [1]{#1}%
\providecommand \enquote  [1]{``#1''}%
\providecommand \bibnamefont  [1]{#1}%
\providecommand \bibfnamefont [1]{#1}%
\providecommand \citenamefont [1]{#1}%
\providecommand \href@noop [0]{\@secondoftwo}%
\providecommand \href [0]{\begingroup \@sanitize@url \@href}%
\providecommand \@href[1]{\@@startlink{#1}\@@href}%
\providecommand \@@href[1]{\endgroup#1\@@endlink}%
\providecommand \@sanitize@url [0]{\catcode `\\12\catcode `\$12\catcode
  `\&12\catcode `\#12\catcode `\^12\catcode `\_12\catcode `\%12\relax}%
\providecommand \@@startlink[1]{}%
\providecommand \@@endlink[0]{}%
\providecommand \url  [0]{\begingroup\@sanitize@url \@url }%
\providecommand \@url [1]{\endgroup\@href {#1}{\urlprefix }}%
\providecommand \urlprefix  [0]{URL }%
\providecommand \Eprint [0]{\href }%
\providecommand \doibase [0]{http://dx.doi.org/}%
\providecommand \selectlanguage [0]{\@gobble}%
\providecommand \bibinfo  [0]{\@secondoftwo}%
\providecommand \bibfield  [0]{\@secondoftwo}%
\providecommand \translation [1]{[#1]}%
\providecommand \BibitemOpen [0]{}%
\providecommand \bibitemStop [0]{}%
\providecommand \bibitemNoStop [0]{.\EOS\space}%
\providecommand \EOS [0]{\spacefactor3000\relax}%
\providecommand \BibitemShut  [1]{\csname bibitem#1\endcsname}%
\let\auto@bib@innerbib\@empty
\bibitem [{\citenamefont {Brown}\ \emph {et~al.}(2013)\citenamefont {Brown},
  \citenamefont {Clark}, \citenamefont {{DuBois}},\ and\ \citenamefont
  {Ceperley}}]{Brown2013}%
  \BibitemOpen
  \bibfield  {author} {\bibinfo {author} {\bibfnamefont {E.~W.}\ \bibnamefont
  {Brown}}, \bibinfo {author} {\bibfnamefont {B.~K.}\ \bibnamefont {Clark}},
  \bibinfo {author} {\bibfnamefont {J.~L.}\ \bibnamefont {{DuBois}}}, \ and\
  \bibinfo {author} {\bibfnamefont {D.~M.}\ \bibnamefont {Ceperley}},\
  }\href@noop {} {\bibfield  {journal} {\bibinfo  {journal} {Phys. Rev. Lett.}\
  }\textbf {\bibinfo {volume} {110}},\ \bibinfo {pages} {146405} (\bibinfo
  {year} {2013})}\BibitemShut {NoStop}%
\bibitem [{\citenamefont {Brown}\ \emph {et~al.}(2014)\citenamefont {Brown},
  \citenamefont {Morales}, \citenamefont {Pierleoni},\ and\ \citenamefont
  {Ceperley}}]{Brown2014}%
  \BibitemOpen
  \bibfield  {author} {\bibinfo {author} {\bibfnamefont {E.}~\bibnamefont
  {Brown}}, \bibinfo {author} {\bibfnamefont {M.~A.}\ \bibnamefont {Morales}},
  \bibinfo {author} {\bibfnamefont {C.}~\bibnamefont {Pierleoni}}, \ and\
  \bibinfo {author} {\bibfnamefont {D.~M.}\ \bibnamefont {Ceperley}},\ }in\
  \href@noop {} {\emph {\bibinfo {booktitle} {Frontiers and Challenges in Warm
  Dense Matter}}},\ \bibinfo {editor} {edited by\ \bibinfo {editor}
  {\bibnamefont {{F. Graziani {\sl et al.}}}}}\ (\bibinfo  {publisher}
  {Springer},\ \bibinfo {year} {2014})\ pp.\ \bibinfo {pages}
  {123--149}\BibitemShut {NoStop}%
\bibitem [{\citenamefont {Dornheim}\ \emph
  {et~al.}(2016{\natexlab{a}})\citenamefont {Dornheim}, \citenamefont {Groth},
  \citenamefont {Sjostrom}, \citenamefont {Malone}, \citenamefont {Foulkes},\
  and\ \citenamefont {Bonitz}}]{Dornheim2016}%
  \BibitemOpen
  \bibfield  {author} {\bibinfo {author} {\bibfnamefont {T.}~\bibnamefont
  {Dornheim}}, \bibinfo {author} {\bibfnamefont {S.}~\bibnamefont {Groth}},
  \bibinfo {author} {\bibfnamefont {T.}~\bibnamefont {Sjostrom}}, \bibinfo
  {author} {\bibfnamefont {F.~D.}\ \bibnamefont {Malone}}, \bibinfo {author}
  {\bibfnamefont {W.~M.~C.}\ \bibnamefont {Foulkes}}, \ and\ \bibinfo {author}
  {\bibfnamefont {M.}~\bibnamefont {Bonitz}},\ }\href@noop {} {\bibfield
  {journal} {\bibinfo  {journal} {Phys. Rev. Lett.}\ }\textbf {\bibinfo
  {volume} {117}},\ \bibinfo {pages} {156403} (\bibinfo {year}
  {2016}{\natexlab{a}})}\BibitemShut {NoStop}%
\bibitem [{\citenamefont {Dornheim}\ \emph
  {et~al.}(2016{\natexlab{b}})\citenamefont {Dornheim}, \citenamefont {Groth},
  \citenamefont {Schoof}, \citenamefont {Hann},\ and\ \citenamefont
  {Bonitz}}]{Dornheim2016b}%
  \BibitemOpen
  \bibfield  {author} {\bibinfo {author} {\bibfnamefont {T.}~\bibnamefont
  {Dornheim}}, \bibinfo {author} {\bibfnamefont {S.}~\bibnamefont {Groth}},
  \bibinfo {author} {\bibfnamefont {T.}~\bibnamefont {Schoof}}, \bibinfo
  {author} {\bibfnamefont {C.}~\bibnamefont {Hann}}, \ and\ \bibinfo {author}
  {\bibfnamefont {M.}~\bibnamefont {Bonitz}},\ }\href@noop {} {\bibfield
  {journal} {\bibinfo  {journal} {Phys. Rev. B}\ }\textbf {\bibinfo {volume}
  {93}},\ \bibinfo {pages} {205134} (\bibinfo {year}
  {2016}{\natexlab{b}})}\BibitemShut {NoStop}%
\bibitem [{\citenamefont {Groth}\ \emph {et~al.}(2016)\citenamefont {Groth},
  \citenamefont {Schoof}, \citenamefont {Dornheim},\ and\ \citenamefont
  {Bonitz}}]{Groth2016}%
  \BibitemOpen
  \bibfield  {author} {\bibinfo {author} {\bibfnamefont {S.}~\bibnamefont
  {Groth}}, \bibinfo {author} {\bibfnamefont {T.}~\bibnamefont {Schoof}},
  \bibinfo {author} {\bibfnamefont {T.}~\bibnamefont {Dornheim}}, \ and\
  \bibinfo {author} {\bibfnamefont {M.}~\bibnamefont {Bonitz}},\ }\href@noop {}
  {\bibfield  {journal} {\bibinfo  {journal} {Phys. Rev. B}\ }\textbf {\bibinfo
  {volume} {93}},\ \bibinfo {pages} {085102} (\bibinfo {year}
  {2016})}\BibitemShut {NoStop}%
\bibitem [{\citenamefont {Groth}\ \emph {et~al.}(2017)\citenamefont {Groth},
  \citenamefont {Dornheim}, \citenamefont {Sjostrom}, \citenamefont {Malone},
  \citenamefont {Foulkes},\ and\ \citenamefont {Bonitz}}]{Groth2017}%
  \BibitemOpen
  \bibfield  {author} {\bibinfo {author} {\bibfnamefont {S.}~\bibnamefont
  {Groth}}, \bibinfo {author} {\bibfnamefont {T.}~\bibnamefont {Dornheim}},
  \bibinfo {author} {\bibfnamefont {T.}~\bibnamefont {Sjostrom}}, \bibinfo
  {author} {\bibfnamefont {F.~D.}\ \bibnamefont {Malone}}, \bibinfo {author}
  {\bibfnamefont {W.~M.~C.}\ \bibnamefont {Foulkes}}, \ and\ \bibinfo {author}
  {\bibfnamefont {M.}~\bibnamefont {Bonitz}},\ }\href@noop {} {\bibfield
  {journal} {\bibinfo  {journal} {Phys. Rev. Lett.}\ }\textbf {\bibinfo
  {volume} {119}},\ \bibinfo {pages} {135001} (\bibinfo {year}
  {2017})}\BibitemShut {NoStop}%
\bibitem [{\citenamefont {Malone}\ \emph {et~al.}(2016)\citenamefont {Malone},
  \citenamefont {Blunt}, \citenamefont {Brown}, \citenamefont {Lee},
  \citenamefont {Spencer}, \citenamefont {Foulkes},\ and\ \citenamefont
  {Shepherd}}]{Malone2016}%
  \BibitemOpen
  \bibfield  {author} {\bibinfo {author} {\bibfnamefont {F.~D.}\ \bibnamefont
  {Malone}}, \bibinfo {author} {\bibfnamefont {N.~S.}\ \bibnamefont {Blunt}},
  \bibinfo {author} {\bibfnamefont {E.~W.}\ \bibnamefont {Brown}}, \bibinfo
  {author} {\bibfnamefont {D.~K.~K.}\ \bibnamefont {Lee}}, \bibinfo {author}
  {\bibfnamefont {J.~S.}\ \bibnamefont {Spencer}}, \bibinfo {author}
  {\bibfnamefont {W.~M.~C.}\ \bibnamefont {Foulkes}}, \ and\ \bibinfo {author}
  {\bibfnamefont {J.~J.}\ \bibnamefont {Shepherd}},\ }\href@noop {} {\bibfield
  {journal} {\bibinfo  {journal} {Phys. Rev. Lett.}\ }\textbf {\bibinfo
  {volume} {117}},\ \bibinfo {pages} {115701} (\bibinfo {year}
  {2016})}\BibitemShut {NoStop}%
\bibitem [{\citenamefont {Filinov}\ \emph {et~al.}(2015)\citenamefont
  {Filinov}, \citenamefont {Fortov}, \citenamefont {Bonitz},\ and\
  \citenamefont {Moldabekov}}]{Filinov2015}%
  \BibitemOpen
  \bibfield  {author} {\bibinfo {author} {\bibfnamefont {V.~S.}\ \bibnamefont
  {Filinov}}, \bibinfo {author} {\bibfnamefont {V.~E.}\ \bibnamefont {Fortov}},
  \bibinfo {author} {\bibfnamefont {M.}~\bibnamefont {Bonitz}}, \ and\ \bibinfo
  {author} {\bibfnamefont {Z.}~\bibnamefont {Moldabekov}},\ }\href@noop {}
  {\bibfield  {journal} {\bibinfo  {journal} {Phys. Rev. E}\ }\textbf {\bibinfo
  {volume} {91}},\ \bibinfo {pages} {033108} (\bibinfo {year}
  {2015})}\BibitemShut {NoStop}%
\bibitem [{\citenamefont {Lerda}(1992)}]{Lerda}%
  \BibitemOpen
  \bibfield  {author} {\bibinfo {author} {\bibfnamefont {A.}~\bibnamefont
  {Lerda}},\ }\href@noop {} {\emph {\bibinfo {title} {Anyons. Quantum mechanics
  of particles with fractional statistics}}},\ Lecture Notes in Physics\
  (\bibinfo  {publisher} {Springer-Verlag},\ \bibinfo {address} {Berlin
  Heidelberg},\ \bibinfo {year} {1992})\BibitemShut {NoStop}%
\bibitem [{\citenamefont {Prieto}(2009)}]{Prieto2009}%
  \BibitemOpen
  \bibfield  {author} {\bibinfo {author} {\bibfnamefont {C.~T.}\ \bibnamefont
  {Prieto}},\ }\href@noop {} {\bibfield  {journal} {\bibinfo  {journal} {J.
  Phys.: Conf. Ser.}\ }\textbf {\bibinfo {volume} {175}},\ \bibinfo {pages}
  {012014} (\bibinfo {year} {2009})}\BibitemShut {NoStop}%
\bibitem [{\citenamefont {Lee}(1999)}]{Lee1999}%
  \BibitemOpen
  \bibfield  {author} {\bibinfo {author} {\bibfnamefont {H.~R.}\ \bibnamefont
  {Lee}},\ }\href@noop {} {\bibfield  {journal} {\bibinfo  {journal} {Journal
  of the Korean Physical Society}\ }\textbf {\bibinfo {volume} {34}},\ \bibinfo
  {pages} {S189} (\bibinfo {year} {1999})}\BibitemShut {NoStop}%
\bibitem [{\citenamefont {Bergeron}\ and\ \citenamefont
  {Semenoff}(1996)}]{Bergeron1996}%
  \BibitemOpen
  \bibfield  {author} {\bibinfo {author} {\bibfnamefont {M.}~\bibnamefont
  {Bergeron}}\ and\ \bibinfo {author} {\bibfnamefont {G.}~\bibnamefont
  {Semenoff}},\ }\href@noop {} {\bibfield  {journal} {\bibinfo  {journal}
  {Annals of Physics}\ }\textbf {\bibinfo {volume} {245}},\ \bibinfo {pages}
  {1} (\bibinfo {year} {1996})}\BibitemShut {NoStop}%
\bibitem [{\citenamefont {Melik-Alaverdian}, \citenamefont {Bonesteel},\ and\
  \citenamefont {Ortiz}(1997)}]{Melik1997}%
  \BibitemOpen
  \bibfield  {author} {\bibinfo {author} {\bibfnamefont {V.}~\bibnamefont
  {Melik-Alaverdian}}, \bibinfo {author} {\bibfnamefont {N.~E.}\ \bibnamefont
  {Bonesteel}}, \ and\ \bibinfo {author} {\bibfnamefont {G.}~\bibnamefont
  {Ortiz}},\ }\href@noop {} {\bibfield  {journal} {\bibinfo  {journal} {Phys.
  Rev. Lett.}\ }\textbf {\bibinfo {volume} {79}},\ \bibinfo {pages} {5286}
  (\bibinfo {year} {1997})}\BibitemShut {NoStop}%
\bibitem [{\citenamefont {Melik-Alaverdian}, \citenamefont {Ortiz},\ and\
  \citenamefont {Bonesteel}(2001)}]{Melik2001}%
  \BibitemOpen
  \bibfield  {author} {\bibinfo {author} {\bibfnamefont {V.}~\bibnamefont
  {Melik-Alaverdian}}, \bibinfo {author} {\bibfnamefont {G.}~\bibnamefont
  {Ortiz}}, \ and\ \bibinfo {author} {\bibfnamefont {N.~E.}\ \bibnamefont
  {Bonesteel}},\ }\href@noop {} {\bibfield  {journal} {\bibinfo  {journal} {J.
  Stat. Phys.}\ }\textbf {\bibinfo {volume} {104}},\ \bibinfo {pages} {449}
  (\bibinfo {year} {2001})}\BibitemShut {NoStop}%
\bibitem [{\citenamefont {Rashid}\ and\ \citenamefont {Yusoff}(2015)}]{Rashid}%
  \BibitemOpen
  \bibinfo {editor} {\bibfnamefont {A.}~\bibnamefont {Rashid}}\ and\ \bibinfo
  {editor} {\bibfnamefont {M.}~\bibnamefont {Yusoff}},\ eds.,\ \href@noop {}
  {\emph {\bibinfo {title} {Graphene-based Energy Devices}}}\ (\bibinfo
  {publisher} {Wiley-VCH},\ \bibinfo {address} {Weinheim},\ \bibinfo {year}
  {2015})\BibitemShut {NoStop}%
\bibitem [{\citenamefont {Tiwari}\ and\ \citenamefont
  {Syv\"aj\"arvi}(2015)}]{Tiwari}%
  \BibitemOpen
  \bibinfo {editor} {\bibfnamefont {A.}~\bibnamefont {Tiwari}}\ and\ \bibinfo
  {editor} {\bibfnamefont {M.}~\bibnamefont {Syv\"aj\"arvi}},\ eds.,\
  \href@noop {} {\emph {\bibinfo {title} {Graphene Materials: Fundamentals and
  Emerging Applications}}}\ (\bibinfo  {publisher} {Scrivener Publishing},\
  \bibinfo {address} {Salem, Massachusetts},\ \bibinfo {year}
  {2015})\BibitemShut {NoStop}%
\bibitem [{\citenamefont {Guo}, \citenamefont {Song},\ and\ \citenamefont
  {Chena}(2010)}]{Guo2010}%
  \BibitemOpen
  \bibfield  {author} {\bibinfo {author} {\bibfnamefont {P.}~\bibnamefont
  {Guo}}, \bibinfo {author} {\bibfnamefont {H.}~\bibnamefont {Song}}, \ and\
  \bibinfo {author} {\bibfnamefont {X.}~\bibnamefont {Chena}},\ }\href@noop {}
  {\bibfield  {journal} {\bibinfo  {journal} {J. Mater. Chem.}\ }\textbf
  {\bibinfo {volume} {20}},\ \bibinfo {pages} {4867} (\bibinfo {year}
  {2010})}\BibitemShut {NoStop}%
\bibitem [{\citenamefont {Cao}\ \emph {et~al.}(2013)\citenamefont {Cao},
  \citenamefont {Wang}, \citenamefont {Xiao}, \citenamefont {Chen},
  \citenamefont {Zhou}, \citenamefont {Ouyang},\ and\ \citenamefont
  {Jia}}]{Cao2013}%
  \BibitemOpen
  \bibfield  {author} {\bibinfo {author} {\bibfnamefont {J.}~\bibnamefont
  {Cao}}, \bibinfo {author} {\bibfnamefont {Y.}~\bibnamefont {Wang}}, \bibinfo
  {author} {\bibfnamefont {P.}~\bibnamefont {Xiao}}, \bibinfo {author}
  {\bibfnamefont {Y.}~\bibnamefont {Chen}}, \bibinfo {author} {\bibfnamefont
  {Y.}~\bibnamefont {Zhou}}, \bibinfo {author} {\bibfnamefont {J.-H.}\
  \bibnamefont {Ouyang}}, \ and\ \bibinfo {author} {\bibfnamefont
  {D.}~\bibnamefont {Jia}},\ }\href@noop {} {\bibfield  {journal} {\bibinfo
  {journal} {Carbon}\ }\textbf {\bibinfo {volume} {56}},\ \bibinfo {pages}
  {389} (\bibinfo {year} {2013})}\BibitemShut {NoStop}%
\bibitem [{\citenamefont {Wu}\ \emph {et~al.}(2013)\citenamefont {Wu},
  \citenamefont {Feng}, \citenamefont {Liu}, \citenamefont {Zhang},\ and\
  \citenamefont {Li}}]{Wu2013}%
  \BibitemOpen
  \bibfield  {author} {\bibinfo {author} {\bibfnamefont {L.}~\bibnamefont
  {Wu}}, \bibinfo {author} {\bibfnamefont {H.}~\bibnamefont {Feng}}, \bibinfo
  {author} {\bibfnamefont {M.}~\bibnamefont {Liu}}, \bibinfo {author}
  {\bibfnamefont {K.}~\bibnamefont {Zhang}}, \ and\ \bibinfo {author}
  {\bibfnamefont {J.}~\bibnamefont {Li}},\ }\href@noop {} {\bibfield  {journal}
  {\bibinfo  {journal} {Nanoscale}\ }\textbf {\bibinfo {volume} {5}},\ \bibinfo
  {pages} {10839} (\bibinfo {year} {2013})}\BibitemShut {NoStop}%
\bibitem [{\citenamefont {Shao}\ \emph {et~al.}(2013)\citenamefont {Shao},
  \citenamefont {Tang}, \citenamefont {Lin}, \citenamefont {Zhang},
  \citenamefont {Yuan}, \citenamefont {Zhang}, \citenamefont {Shinyaa},\ and\
  \citenamefont {Qinc}}]{Shao2013}%
  \BibitemOpen
  \bibfield  {author} {\bibinfo {author} {\bibfnamefont {Q.}~\bibnamefont
  {Shao}}, \bibinfo {author} {\bibfnamefont {J.}~\bibnamefont {Tang}}, \bibinfo
  {author} {\bibfnamefont {Y.}~\bibnamefont {Lin}}, \bibinfo {author}
  {\bibfnamefont {F.}~\bibnamefont {Zhang}}, \bibinfo {author} {\bibfnamefont
  {J.}~\bibnamefont {Yuan}}, \bibinfo {author} {\bibfnamefont {H.}~\bibnamefont
  {Zhang}}, \bibinfo {author} {\bibfnamefont {N.}~\bibnamefont {Shinyaa}}, \
  and\ \bibinfo {author} {\bibfnamefont {L.-C.}\ \bibnamefont {Qinc}},\
  }\href@noop {} {\bibfield  {journal} {\bibinfo  {journal} {J. Mater. Chem.
  A}\ }\textbf {\bibinfo {volume} {1}},\ \bibinfo {pages} {15423} (\bibinfo
  {year} {2013})}\BibitemShut {NoStop}%
\bibitem [{\citenamefont {Zhao}, \citenamefont {Chen},\ and\ \citenamefont
  {Wu}(2016)}]{Zhao2016}%
  \BibitemOpen
  \bibfield  {author} {\bibinfo {author} {\bibfnamefont {Y.}~\bibnamefont
  {Zhao}}, \bibinfo {author} {\bibfnamefont {M.}~\bibnamefont {Chen}}, \ and\
  \bibinfo {author} {\bibfnamefont {L.}~\bibnamefont {Wu}},\ }\href@noop {}
  {\bibfield  {journal} {\bibinfo  {journal} {Nanotechnology}\ }\textbf
  {\bibinfo {volume} {27}},\ \bibinfo {pages} {342001} (\bibinfo {year}
  {2016})}\BibitemShut {NoStop}%
\bibitem [{\citenamefont {Cho}, \citenamefont {Lee},\ and\ \citenamefont
  {Y.C.Kang}(2016)}]{Cho2016}%
  \BibitemOpen
  \bibfield  {author} {\bibinfo {author} {\bibfnamefont {J.~S.}\ \bibnamefont
  {Cho}}, \bibinfo {author} {\bibfnamefont {J.-K.}\ \bibnamefont {Lee}}, \ and\
  \bibinfo {author} {\bibnamefont {Y.C.Kang}},\ }\href@noop {} {\bibfield
  {journal} {\bibinfo  {journal} {Scientific Reports}\ }\textbf {\bibinfo
  {volume} {6}} (\bibinfo {year} {2016})}\BibitemShut {NoStop}%
\bibitem [{\citenamefont {Hao}\ \emph {et~al.}(2016)\citenamefont {Hao},
  \citenamefont {Xuefen}, \citenamefont {Liangdong},\ and\ \citenamefont
  {Xiaohui}}]{Hao2016}%
  \BibitemOpen
  \bibfield  {author} {\bibinfo {author} {\bibfnamefont {D.}~\bibnamefont
  {Hao}}, \bibinfo {author} {\bibfnamefont {C.}~\bibnamefont {Xuefen}},
  \bibinfo {author} {\bibfnamefont {Q.}~\bibnamefont {Liangdong}}, \ and\
  \bibinfo {author} {\bibfnamefont {Z.}~\bibnamefont {Xiaohui}},\ }\href@noop
  {} {\bibfield  {journal} {\bibinfo  {journal} {Rare Metal Materials and
  Engineering}\ }\textbf {\bibinfo {volume} {45}},\ \bibinfo {pages} {1669}
  (\bibinfo {year} {2016})}\BibitemShut {NoStop}%
\bibitem [{\citenamefont {Huang}\ \emph {et~al.}(2017)\citenamefont {Huang},
  \citenamefont {Ding}, \citenamefont {Chen}, \citenamefont {Hao},
  \citenamefont {Lai}, \citenamefont {Peng}, \citenamefont {Tu}, \citenamefont
  {Cao},\ and\ \citenamefont {Li}}]{Huang2017}%
  \BibitemOpen
  \bibfield  {author} {\bibinfo {author} {\bibfnamefont {W.}~\bibnamefont
  {Huang}}, \bibinfo {author} {\bibfnamefont {S.}~\bibnamefont {Ding}},
  \bibinfo {author} {\bibfnamefont {Y.}~\bibnamefont {Chen}}, \bibinfo {author}
  {\bibfnamefont {W.}~\bibnamefont {Hao}}, \bibinfo {author} {\bibfnamefont
  {X.}~\bibnamefont {Lai}}, \bibinfo {author} {\bibfnamefont {J.}~\bibnamefont
  {Peng}}, \bibinfo {author} {\bibfnamefont {J.}~\bibnamefont {Tu}}, \bibinfo
  {author} {\bibfnamefont {Y.}~\bibnamefont {Cao}}, \ and\ \bibinfo {author}
  {\bibfnamefont {X.}~\bibnamefont {Li}},\ }\href@noop {} {\bibfield  {journal}
  {\bibinfo  {journal} {Scientific Reports}\ }\textbf {\bibinfo {volume} {7}}
  (\bibinfo {year} {2017})}\BibitemShut {NoStop}%
\bibitem [{\citenamefont {Bi}\ \emph {et~al.}(2017)\citenamefont {Bi},
  \citenamefont {Chen}, \citenamefont {Yang}, \citenamefont {Ye}, \citenamefont
  {Yin},\ and\ \citenamefont {Han}}]{Chen2017}%
  \BibitemOpen
  \bibfield  {author} {\bibinfo {author} {\bibfnamefont {E.}~\bibnamefont
  {Bi}}, \bibinfo {author} {\bibfnamefont {H.}~\bibnamefont {Chen}}, \bibinfo
  {author} {\bibfnamefont {X.}~\bibnamefont {Yang}}, \bibinfo {author}
  {\bibfnamefont {F.}~\bibnamefont {Ye}}, \bibinfo {author} {\bibfnamefont
  {M.}~\bibnamefont {Yin}}, \ and\ \bibinfo {author} {\bibfnamefont
  {L.}~\bibnamefont {Han}},\ }\href@noop {} {\bibfield  {journal} {\bibinfo
  {journal} {Scientific Reports}\ }\textbf {\bibinfo {volume} {5}} (\bibinfo
  {year} {2017})}\BibitemShut {NoStop}%
\bibitem [{\citenamefont {Fantoni}(2012{\natexlab{a}})}]{Fantoni12e}%
  \BibitemOpen
  \bibfield  {author} {\bibinfo {author} {\bibfnamefont {R.}~\bibnamefont
  {Fantoni}},\ }\href {\doibase 10.1088/1742-5468/2012/10/P10024} {\bibfield
  {journal} {\bibinfo  {journal} {J. Stat. Mech.}\ ,\ \bibinfo {pages}
  {P10024}} (\bibinfo {year} {2012}{\natexlab{a}})}\BibitemShut {NoStop}%
\bibitem [{\citenamefont {Schulman}(1981)}]{Schulman}%
  \BibitemOpen
  \bibfield  {author} {\bibinfo {author} {\bibfnamefont {L.~S.}\ \bibnamefont
  {Schulman}},\ }\href@noop {} {\emph {\bibinfo {title} {Techniques and
  applications of path integrals}}}\ (\bibinfo  {publisher} {John Wiley \&
  Sons},\ \bibinfo {year} {1981})\ \bibinfo {note} {chapter 24}\BibitemShut
  {NoStop}%
\bibitem [{\citenamefont {{D. M. Ceperley}}(1995)}]{Ceperley1995}%
  \BibitemOpen
  \bibfield  {author} {\bibinfo {author} {\bibnamefont {{D. M. Ceperley}}},\
  }\href@noop {} {\bibfield  {journal} {\bibinfo  {journal} {Rev. Mod. Phys.}\
  }\textbf {\bibinfo {volume} {67}},\ \bibinfo {pages} {279} (\bibinfo {year}
  {1995})}\BibitemShut {NoStop}%
\bibitem [{\citenamefont {Ceperley}(1991)}]{Ceperley1991}%
  \BibitemOpen
  \bibfield  {author} {\bibinfo {author} {\bibfnamefont {D.~M.}\ \bibnamefont
  {Ceperley}},\ }\href@noop {} {\bibfield  {journal} {\bibinfo  {journal} {J.
  Stat. Phys.}\ }\textbf {\bibinfo {volume} {63}},\ \bibinfo {pages} {1237}
  (\bibinfo {year} {1991})}\BibitemShut {NoStop}%
\bibitem [{\citenamefont {Ceperley}(1996)}]{Ceperley1996}%
  \BibitemOpen
  \bibfield  {author} {\bibinfo {author} {\bibfnamefont {D.~M.}\ \bibnamefont
  {Ceperley}},\ }in\ \href@noop {} {\emph {\bibinfo {booktitle} {Monte Carlo
  and Molecular Dynamics of Condensed Matter Systems}}},\ \bibinfo {editor}
  {edited by\ \bibinfo {editor} {\bibfnamefont {K.}~\bibnamefont {Binder}}\
  and\ \bibinfo {editor} {\bibfnamefont {G.}~\bibnamefont {Ciccotti}}}\
  (\bibinfo  {publisher} {Editrice Compositori},\ \bibinfo {address} {Bologna,
  Italy},\ \bibinfo {year} {1996})\BibitemShut {NoStop}%
\bibitem [{\citenamefont {Prokof'ev}, \citenamefont {Svistunov},\ and\
  \citenamefont {Tupitsyn}(1998)}]{Prokofev1998}%
  \BibitemOpen
  \bibfield  {author} {\bibinfo {author} {\bibfnamefont {N.~V.}\ \bibnamefont
  {Prokof'ev}}, \bibinfo {author} {\bibfnamefont {B.~V.}\ \bibnamefont
  {Svistunov}}, \ and\ \bibinfo {author} {\bibfnamefont {I.~S.}\ \bibnamefont
  {Tupitsyn}},\ }\href@noop {} {\bibfield  {journal} {\bibinfo  {journal} {J.
  Exp. Theor. Phys.}\ }\textbf {\bibinfo {volume} {87}},\ \bibinfo {pages}
  {310} (\bibinfo {year} {1998})}\BibitemShut {NoStop}%
\bibitem [{\citenamefont {Boninsegni}, \citenamefont {Prokof'ev},\ and\
  \citenamefont {Svistunov}(2006{\natexlab{a}})}]{Boninsegni2006a}%
  \BibitemOpen
  \bibfield  {author} {\bibinfo {author} {\bibfnamefont {M.}~\bibnamefont
  {Boninsegni}}, \bibinfo {author} {\bibfnamefont {N.}~\bibnamefont
  {Prokof'ev}}, \ and\ \bibinfo {author} {\bibfnamefont {B.}~\bibnamefont
  {Svistunov}},\ }\href@noop {} {\bibfield  {journal} {\bibinfo  {journal}
  {Phys. Rev. Lett.}\ }\textbf {\bibinfo {volume} {96}},\ \bibinfo {pages}
  {070601} (\bibinfo {year} {2006}{\natexlab{a}})}\BibitemShut {NoStop}%
\bibitem [{\citenamefont {Boninsegni}, \citenamefont {Prokof'ev},\ and\
  \citenamefont {Svistunov}(2006{\natexlab{b}})}]{Boninsegni2006b}%
  \BibitemOpen
  \bibfield  {author} {\bibinfo {author} {\bibfnamefont {M.}~\bibnamefont
  {Boninsegni}}, \bibinfo {author} {\bibfnamefont {N.~V.}\ \bibnamefont
  {Prokof'ev}}, \ and\ \bibinfo {author} {\bibfnamefont {B.~V.}\ \bibnamefont
  {Svistunov}},\ }\href@noop {} {\bibfield  {journal} {\bibinfo  {journal}
  {Phys. Rev. E}\ }\textbf {\bibinfo {volume} {74}},\ \bibinfo {pages} {036701}
  (\bibinfo {year} {2006}{\natexlab{b}})}\BibitemShut {NoStop}%
\bibitem [{\citenamefont {Fantoni}(2018{\natexlab{a}})}]{Fantoni2018a}%
  \BibitemOpen
  \bibfield  {author} {\bibinfo {author} {\bibfnamefont {R.}~\bibnamefont
  {Fantoni}},\ }\href@noop {} {\bibfield  {journal} {\bibinfo  {journal} {Eur.
  Phys. J. B}\ } (\bibinfo {year} {2018}{\natexlab{a}})},\ \bibinfo {note} {in
  preparation}\BibitemShut {NoStop}%
\bibitem [{\citenamefont {Fantoni}\ and\ \citenamefont
  {Moroni}(2014)}]{Fantoni14d}%
  \BibitemOpen
  \bibfield  {author} {\bibinfo {author} {\bibfnamefont {R.}~\bibnamefont
  {Fantoni}}\ and\ \bibinfo {author} {\bibfnamefont {S.}~\bibnamefont
  {Moroni}},\ }\href {\doibase 10.1063/1.4895974} {\bibfield  {journal}
  {\bibinfo  {journal} {J. Chem. Phys.}\ }\textbf {\bibinfo {volume} {141}},\
  \bibinfo {pages} {114110} (\bibinfo {year} {2014})}\BibitemShut {NoStop}%
\bibitem [{\citenamefont {Los}, \citenamefont {andM. I.~Katsnelson},\ and\
  \citenamefont {Fasolino}(2015)}]{Los2015}%
  \BibitemOpen
  \bibfield  {author} {\bibinfo {author} {\bibfnamefont {J.~H.}\ \bibnamefont
  {Los}}, \bibinfo {author} {\bibfnamefont {K.~V.~Z.}\ \bibnamefont {andM.
  I.~Katsnelson}}, \ and\ \bibinfo {author} {\bibfnamefont {A.}~\bibnamefont
  {Fasolino}},\ }\href@noop {} {\bibfield  {journal} {\bibinfo  {journal}
  {Phys. Rev. B}\ }\textbf {\bibinfo {volume} {91}},\ \bibinfo {pages} {045415}
  (\bibinfo {year} {2015})}\BibitemShut {NoStop}%
\bibitem [{\citenamefont {Fantoni}, \citenamefont {Jancovici},\ and\
  \citenamefont {T\'{e}llez}(2003)}]{Fantoni03jsp}%
  \BibitemOpen
  \bibfield  {author} {\bibinfo {author} {\bibfnamefont {R.}~\bibnamefont
  {Fantoni}}, \bibinfo {author} {\bibfnamefont {B.}~\bibnamefont {Jancovici}},
  \ and\ \bibinfo {author} {\bibfnamefont {G.}~\bibnamefont {T\'{e}llez}},\
  }\href@noop {} {\bibfield  {journal} {\bibinfo  {journal} {J. Stat. Phys.}\
  }\textbf {\bibinfo {volume} {{\bf 112}}},\ \bibinfo {pages} {27} (\bibinfo
  {year} {2003})}\BibitemShut {NoStop}%
\bibitem [{\citenamefont {Fantoni}\ and\ \citenamefont
  {T\'ellez}(2008)}]{Fantoni2008}%
  \BibitemOpen
  \bibfield  {author} {\bibinfo {author} {\bibfnamefont {R.}~\bibnamefont
  {Fantoni}}\ and\ \bibinfo {author} {\bibfnamefont {G.}~\bibnamefont
  {T\'ellez}},\ }\href@noop {} {\bibfield  {journal} {\bibinfo  {journal} {J.
  Stat. Phys.}\ }\textbf {\bibinfo {volume} {133}},\ \bibinfo {pages} {449}
  (\bibinfo {year} {2008})}\BibitemShut {NoStop}%
\bibitem [{\citenamefont {Fantoni}(2012{\natexlab{b}})}]{Fantoni2012}%
  \BibitemOpen
  \bibfield  {author} {\bibinfo {author} {\bibfnamefont {R.}~\bibnamefont
  {Fantoni}},\ }\href@noop {} {\bibfield  {journal} {\bibinfo  {journal} {J.
  Stat. Mech.}\ ,\ \bibinfo {pages} {P04015}} (\bibinfo {year}
  {2012}{\natexlab{b}})}\BibitemShut {NoStop}%
\bibitem [{\citenamefont {Fantoni}(2012{\natexlab{c}})}]{Fantoni2012b}%
  \BibitemOpen
  \bibfield  {author} {\bibinfo {author} {\bibfnamefont {R.}~\bibnamefont
  {Fantoni}},\ }\href@noop {} {\bibfield  {journal} {\bibinfo  {journal} {J.
  Stat. Mech.}\ ,\ \bibinfo {pages} {P10024}} (\bibinfo {year}
  {2012}{\natexlab{c}})}\BibitemShut {NoStop}%
\bibitem [{\citenamefont {Fantoni}(2016)}]{Fantoni2016}%
  \BibitemOpen
  \bibfield  {author} {\bibinfo {author} {\bibfnamefont {R.}~\bibnamefont
  {Fantoni}},\ }\href@noop {} {\bibfield  {journal} {\bibinfo  {journal} {J.
  Stat. Phys.}\ }\textbf {\bibinfo {volume} {163}},\ \bibinfo {pages} {1247}
  (\bibinfo {year} {2016})}\BibitemShut {NoStop}%
\bibitem [{\citenamefont {Fantoni}(2017)}]{Fantoni17b}%
  \BibitemOpen
  \bibfield  {author} {\bibinfo {author} {\bibfnamefont {R.}~\bibnamefont
  {Fantoni}},\ }\href {\doibase 10.1016/j.physa.2017.02.064} {\bibfield
  {journal} {\bibinfo  {journal} {Physica A}\ }\textbf {\bibinfo {volume}
  {477C}},\ \bibinfo {pages} {187} (\bibinfo {year} {2017})}\BibitemShut
  {NoStop}%
\bibitem [{\citenamefont {Fantoni}, \citenamefont {Salari},\ and\ \citenamefont
  {Klumperman}(2012)}]{Fantoni12c}%
  \BibitemOpen
  \bibfield  {author} {\bibinfo {author} {\bibfnamefont {R.}~\bibnamefont
  {Fantoni}}, \bibinfo {author} {\bibfnamefont {J.~W.~O.}\ \bibnamefont
  {Salari}}, \ and\ \bibinfo {author} {\bibfnamefont {B.}~\bibnamefont
  {Klumperman}},\ }\href {\doibase 10.1103/PhysRevE.85.061404} {\bibfield
  {journal} {\bibinfo  {journal} {Phys. Rev. E}\ }\textbf {\bibinfo {volume}
  {85}},\ \bibinfo {pages} {061404} (\bibinfo {year} {2012})}\BibitemShut
  {NoStop}%
\bibitem [{\citenamefont {Fantoni}(2018{\natexlab{b}})}]{Fantoni2018b}%
  \BibitemOpen
  \bibfield  {author} {\bibinfo {author} {\bibfnamefont {R.}~\bibnamefont
  {Fantoni}},\ }\href@noop {} {\bibfield  {journal} {\bibinfo  {journal} {Int.
  J. Mod. Phys. C}\ }\textbf {\bibinfo {volume} {29}},\ \bibinfo {pages}
  {1850028} (\bibinfo {year} {2018}{\natexlab{b}})}\BibitemShut {NoStop}%
\end{thebibliography}%

\end{document}